\documentclass[fleqn,usenatbib]{mnras}

\usepackage{mathptmx}

\usepackage[T1]{fontenc}

\DeclareRobustCommand{\VAN}[3]{#2}
\let\VANthebibliography\thebibliography
\def\thebibliography{\DeclareRobustCommand{\VAN}[3]{##3}\VANthebibliography}

\usepackage{graphicx}	
\usepackage{amsmath}	
\usepackage{amssymb}	
\usepackage{txfonts}
\usepackage{hyperref}
\usepackage{natbib}
\usepackage[singlelinecheck=false,justification=justified]{caption}
\usepackage{array,boldline}
\usepackage{booktabs}
\usepackage{tabularx}
\usepackage[scientific-notation=true]{siunitx}
\usepackage{longtable}
\usepackage{float}
\usepackage[utf8]{inputenc}

\title[ETVs in PCEBs an update on recent models]{Eclipse timing variations in 7 post-common envelope binaries: an update and analysis of their recently proposed circumbinary models}

\author[D. Pulley et al.]{
D. Pulley,$^{1}$\thanks{E-mail: david@davidpulley.co.uk}
I. D. Sharp,$^{1}$\thanks{E-mail: ian.sharp@astro-sharp.com}
J. Mallett$^{1}$\thanks{E-mail: john@astro.me.uk}
and S. von Harrach$^{1}$
\\
$^{1}$British Astronomical Association,  Ossington Chambers, 6/8 Castle Gate, Newark, NG24 1AX, UK\\
}

\date{Accepted XXX. Received YYY; in original form ZZZ}

\pubyear{2022}

\begin{document}
\label{firstpage}
\pagerange{\pageref{firstpage}--\pageref{lastpage}}
\maketitle

\begin{abstract}
{Eclipsing, short-period post-common envelope binaries (PCEBs) have been studied for several decades by eclipse timing variations (ETVs) which have been interpreted as being caused by circumbinary bodies. In this paper we report 355 new observations of 7 PECBs (HS0705+6700, NN Ser, NSVS 07826147, NSVS 14256825, NY Vir, QS Vir and RR Cae) and examine how the recent proposed models of these systems compare with our new observations. We find that none of the 18 recent models fit accurately with our data. We review alternative mechanisms of the ETVs, including magnetic effects, but conclude that they do not predict our results. Although we cannot exclude the presence of circumbinary bodies a combination of several mechanisms may be required to explain the observed ETVs.}
\end{abstract}

\begin{keywords}
binaries: close --
binaries: eclipsing --
planetary systems --
subdwarfs --
white dwarfs --
stars: individual: HS0705+6700, NN Ser, NSVS 07826147, NSVS 14256825, NY Vir, QS Vir, RR Cae
\end{keywords}


\section{Introduction}
It has been over three decades since the discovery of the first two exoplanets orbiting the pulsar PSR B1257+12 \citep{wolszczan1992planetary}.  Since then more than 5900 exoplanets/brown dwarfs (BD) have been recorded encircling stellar objects of which only 540, less than 10\%, are found orbiting multi-stellar systems with the radial velocity and transit discovery methods accounting for 494 of these.  The remaining 46 exoplanet/BDs were discovered by astrometry (1), pulsar timing (1), transit time variations (2), microlensing (10), imaging (15) and eclipse timing variations (17)\footnote[2]{Data obtained from the NASA Exoplanet Archive (NEA) as of 2025 June, \url{https://exoplanetarchive.ipac.caltech.edu/docs/counts_detail.html}}.

Whilst ETVs have provided only 3.1\% of confirmed sub-stellar objects orbiting multi-stellar systems, they have been extensively researched over the past twenty years as their eclipse times can be determined with high precision.  This is particularly so for short period, $\mathrm{\sim}$2h, post common envelope binary systems where the very hot subdwarf (sdB) or white dwarf (WD) primary is separated from its cool dwarf M (dM) main sequence star by typically less than 10$R_{\sun}$.  ETVs can occur when a third body, and in this case an exoplanet/BD, orbits the binary pair causing the centre of mass of the binary pair to move relative to the observer.  This change in distance of the binary pair from the observer will lead to a light travel time effect (LTTE) causing eclipses to occur earlier or later than for the binary system alone. This effect can then be monitored over time by recording the difference between the observed and calculated (O - C) eclipse times. Reviews of the ETV approach applied to these systems can be found in \cite{zorotovic2013origin} and \cite{lohr2014period}.

The first claimed sub-stellar mass object discovered from ETV observations, a brown dwarf, was reported to orbit the PCEB V471 Tau \citep{guinan2000best} but, in 2014, direct imaging by the Very Large Telescope SPHERE planet finder resulted in a non-detection.  Since 2001, there have been many claimed circumbinary planets/BDs orbiting eclipsing binaries and detected by monitoring their ETVs.  Although not a comprehensive list, these include DE CVn, DP Leo, HS2231+2441, HU Aqr, Kepler-451, Kepler-1600, NSVS 01286630, UZ For and V2134 Oph together with those listed in Table \ref{table_objects_summary}.  Over the past two decades more than 50 circumbinary models have been listed for these 16 eclipsing binaries with some, e.g. NY Vir and HS0705+6700, having been significantly revised many times as new data was published.

Historically it is found, and particularly so for PCEBs, that circumbinary models have been unable to predict eclipse times into the future by more than a few years (\citeauthor{pulley2018quest} \citeyear{pulley2018quest};  \citeauthor{pulley2022eclipse} \citeyear{pulley2022eclipse}) and other mechanisms for generating these ETVs have been considered.  These alternative mechanisms are many and here we have assumed that the underlying ephemeris can be represented in quadratic form where the quadratic coefficient may be positive, negative or zero. This generalised quadratic ephemeris suggests the presence of secular period changes which are real (or intrinsic) changes within the binary system and most commonly include mechanisms such as (i) mass transfer (ii) mass loss (iii) gravitational effects from the third body and (iv) angular momentum loss (AML) through stellar winds or gravitational radiation.  The very short binary period and detached nature of these PCEB systems suggest that mechanisms (i), (ii) and (iii) do not contribute significantly to ETVs in these systems.

Other possible causes of ETVs include quasi-cyclical mechanisms where (i) apsidal motion (ii) LTTE from circumbinary bodies or (iii) magnetic effects within the secondary star e.g. \cite{applegate1992mechanism} and \cite{lanza2020internal}, are the main contenders.  Apsidal motion is considered unlikely in short period PCEBs since the binary orbit is expected to circularise very quickly.  An overview of these mechanisms is given in Section 2 of \cite{borkovits2025then} where they include a "spurious" category e.g. ETVs caused by star spots and stellar oscillations.  From the perspective of PCEBs, the most commonly referred to mechanisms of ETVs are AML, LTTE from circumbinary companions and magnetic effects. 

In this paper we review 18 circumbinary models associated with the seven eclipsing binaries listed in Table \ref{table_objects_summary} which, apart from QS Vir, date from 2021. In Section 2  we outline our observations and data reduction techniques we employed and in Section 3 consider these models in light of our new observations. We discuss our findings for these systems in Section 4 and in Section 5 review the ETV methodology for identifying circumbinary planets. 

\section{Observations and data reduction}
We report 355 new times of minima of seven sdB and WD PCEB systems observed between 2017 September and 2025 June using the filters and telescopes listed in the two data tables which are available as on-line supplementary material. A small sample of our data is shown in Appendix A, Table \ref{table_A1} and Table \ref{table_A2} lists the telescopes used. Table \ref{table_objects_summary} summarises our observations by star.

The objective of our image acquisition and data reduction processes is to accurately determine the times of minima using the light curves which result from differential aperture photometry, so it is more important to achieve a high temporal resolution than to derive the precise magnitude of the binary system.  To maximize the signal-to-noise ratio with the shortest possible exposure time, many observations were made with either no or a broad-band filter and in these cases the V-band magnitudes of the comparison stars were used. Where filters were employed, the appropriate filter band magnitudes were used.

The comparison stars’ catalogue magnitudes for the various filters were taken from the American Association of Variable Star Observers Photometric All Sky Survey (APASS) catalogue or, in some cases, the US Naval Observatory CCD Astrograph Catalog, UCAC4, and were chosen to be similar to the target magnitudes. As far as possible, comparison stars with similar colour indices to the target stars were selected.

The effects of differing atmospheric extinctions were minimized by making all observations at altitudes greater than 40\textdegree\ and where the Moon’s effulgence would not cause a light gradient across the image frames. Checks were also made for other potential causes of measurement error including star roundness to ensure effective guiding and the images were free from aircraft and satellite trails.

All images were calibrated using bias, dark and flat frames and then analysed with either AstroArt\footnote[3]{AstroArt, \url{http://www.msb-astroart.com/}} or MaxIm DL\footnote[4]{MaxIm\ DL, \url{http://www.diffractionlimited.com/}} software packages. The source flux was determined with aperture photometry using a variable aperture, whereby the radius is scaled according to the  full width at half-maximum of the target image. Variations in observing conditions were accounted for by determining the flux relative to comparison stars in the field of view.

During the processing of the images the signal-to-noise ratio of all measurements and the deviation of the reference stars from the reference magnitude were also checked for each measurement. Finally, the check star reference magnitude was confirmed to be within a tolerance range of the reference magnitude.

All of our new timings were first converted to barycentric Julian date dynamical time (BJD\_TBD) using either the time utilities of the Ohio State University\footnote[5]{Ohio State University, \url{http://astroutils.astronomy.ohio-state.edu/time/}} or custom-written Python code using the Astropy\footnote[6]{Astropy, \url{https://www.astropy.org/}} library. For most data we calculated the times of minima using the procedure of \cite{kwee1956method} coded in our own Python utilities.  For those systems with a WD primary and exhibiting a flat bottom to the eclipse curve, we derived the best fit equations to the ingress and egress with linear regression before computing the eclipse midpoint using our own Python software utilities to standardise this process.

Our new timings were combined with previously published times of minima and, where appropriate, the historic times were converted to BJD\_TBD before computing new linear and/or quadratic ephemerides and O - C residuals. 

The results of the processing are uploaded to our proprietary website where they can be peer-reviewed and checked at any time by our team members. Finally, every FIT image file is archived in case it is necessary to re-process the images.

\begin{table*}
\caption{Summary of the seven PCEBs reported in this paper including 355 new times of primary minima recorded.  No secondary minima have been observed.}
\label{table_objects_summary}
\begin{tabular}{p{2.75cm} p{1.75cm} p{1.75cm} l c c l}
\toprule
Object & RA\newline(J2000) & Dec\newline(J2000) & Period (d) & Mag. & Primary Minima & Observing Period\\
\midrule
HS0705+6700 & 07 10 42.05 & +66 55 43.52 & 0.095646620 & 14.60 (G) & 112 & 2022 Feb - 2025 Jun\\
NN Ser & 15 52 56.12 & +12 54 44.43 & 0.130080080 & 16.51 (V) & 20 & 2022 Feb - 2025 Jun\\
NSVS 07826147 & 15 33 49.44 & +37 59 28.10 & 0.161770447 & 13.08 (V) & 51 & 2017 Sep - 2025 Jun\\
NSVS 14256825 & 20 20 00.46 & +04 37 56.52 & 0.1103741092 & 13.34 (R) & 56 & 2022 May - 2025 Jun\\
NY Vir & 13 38 48.15 & -02 01 49.21 & 0.1010159602 & 13.66 (V) & 47 & 2022 May - 2025 Jun\\
QS Vir & 13 49 52.00 & -13 13 37.00 & 0.150757442 & 14.40 (V) & 34 & 2020 Jan - 2025 Jun\\
RR Cae & 04 21 05.56 & -48 39 07.06 & 00.303703648 & 14.40 (V) & 35 & 2022 Aug - 2025 Apr\\
\bottomrule
\end{tabular}
\end{table*}

\section{New Planetary Models}
In this section the recently published circumbinary models for each of the 7 PCEB systems are examined and updated with our latest minima timings.

\subsection{Transiting Exoplanet Survey Satellite (TESS) Data} \label{ssec:tess}
It should be noted that 3 of the papers summarised below (\cite{rattanamala2023eclipse}, \cite{Er_Özdönmez_Nasiroglu_Kenger_2024} and \cite{zervas2024nsvs}) include eclipse timings obtained from the TESS\footnote[7]{TESS, \url{https://tess.mit.edu/}}.

Most studies cover data collected over several decades and the TESS data shows up as highly scattered, short bursts which have little effect on the computation of any particular ephemeris but, summing the squares of the deviations could lead to the TESS uncertainties dominating the ephemeris uncertainties.

It is possible that the relatively long integration time of the TESS images, which is dominated by images stacked every 2 minutes, could contribute to the high degree scatter compared to the ground-based results but, whatever the reason for this scatter, we have chosen not to include TESS timings in our ephemeris calculations.

\subsection{HS0705+6700 (V470 Cam)}
HS0705+6700 was observed as part of the Hamburg Schmidt Quasar Survey and subsequently identified as a short period, $\mathrm{\sim}$2.3h, eclipsing PCEB comprising a 14.7 magnitude sdB primary with an M dwarf companion \citep{drechsel2001hs}.  

Since its discovery this system has been extensively studied and between 2009 and 2022 eight circumbinary models have been proposed to explain the apparent quasi-periodic variation observed in its eclipse timings.  However seven of these models failed to predict the system’s long-term behaviour within twelve months of their publication.  The eighth model could not be tested as not all of its orbital parameters were specified.  A review of these models, together with the binary system parameters, can be found in \cite{pulley2022eclipse}.

\begin{figure}
	\captionsetup{width=8.50cm}
   \includegraphics[width=8.50cm]{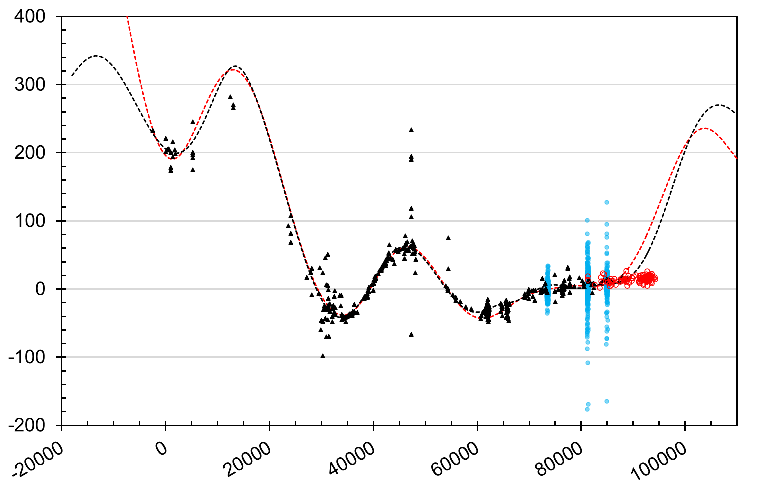}
     \caption{O – C chart for HS0705+6700 showing the two models from \protect\cite{Er_Özdönmez_Nasiroglu_Kenger_2024} updated with new data from this paper. Historical data $\mathrm{\sim}$ black filled triangles; TESS data $\mathrm{\sim}$ blue filled circles; Our new data $\mathrm{\sim}$ red open circles; Quadratic with 2 bodies model $\mathrm{\sim}$ red dashed line; Linear with 3 bodies model $\mathrm{\sim}$ black dashed line. Vertical axis is seconds and horizontal axis cycle number.}
     \label{HS0705_Er_Both}
\end{figure}

The most recent study, \cite{Er_Özdönmez_Nasiroglu_Kenger_2024}, added 88 new times of minima and a further 5 binned data points from TESS (see Section \ref{ssec:tess}) through to 2024 January. With all available data they constructed two circumbinary models with (i) a linear ephemeris and three circumbinary brown dwarfs and (ii) a quadratic ephemeris with two circumbinary brown dwarfs.  Their analysis recognised that both models were unstable on a timescale of less than 2000y whilst our recent data also indicates that their two circumbinary models would fail to predict future eclipse times correctly, see Fig. \ref{HS0705_Er_Both}.

\begin{figure}
	\captionsetup{width=8.50cm}
   \includegraphics[width=8.50cm]{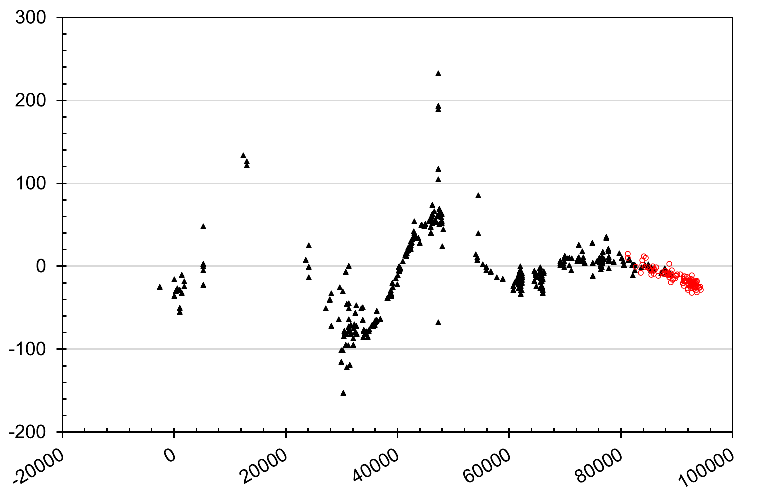}
     \caption{O – C chart for HS0705+6700 showing the quadratic ephemeris of Eq. 1. Historical data $\mathrm{\sim}$ black filled triangles; Our new data $\mathrm{\sim}$ red open circles. Vertical axis is seconds and horizontal axis cycle number.}
     \label{HS0705_Quad_Ephem_graph}
\end{figure}

We present a further 112 new times of primary minima post 2022 February and compute the best fit quadratic ephemeris, Eq. \ref{HS0705_ephem_quad}, which excludes all TESS data:
   \begin{equation}\label{HS0705_ephem_quad}
   \begin{aligned}
      BTDB_{\mathrm{min}} ={} & 2451822.76092(8) + 0.095646620(3)E  + \\
      & 7.2(3)  \times 10^{-13}E^2
   \end{aligned}
   \end{equation}
Where E is the cycle number. The O – C residuals calculated from Eq. \ref{HS0705_ephem_quad} are shown in Fig. \ref{HS0705_Quad_Ephem_graph}.  This chart shows data at E $\mathrm{\sim}$5200 and $\mathrm{\sim}$47308 extracted from IBVS 6029, 6063 and 6153. As this data exhibits an unexplained wide spread, e.g. $\mathrm{\sim}$300s over 1.5 cycles at E $\mathrm{\sim}$47308, we have not included it in the derivation of Eq. \ref{HS0705_ephem_quad}.


\subsection{NN Ser}
NN Ser is a short period, $\mathrm{\sim}$3.1h, PCEB system comprising a hot, 57,000K, WD and an dwarf M4 companion.  The compact nature of the WD results in a primary eclipse with rapid, $\mathrm{\sim}$80s, ingress and egress transitions, eclipse duration of some 600s, and an associated V band brightness decrease of approximately 5 magnitudes. See \cite{parsons2010precise} for further system parameters.

Times of eclipse minima have been recorded since 1989, \cite{1989A&A...213L..15H}, and prior to 2023 five circumbinary models have been proposed to explain the observed variations in eclipse times.  However none of these models have been able to predict the long-term future of this system with most failing within a few months of publication, see \cite{pulley2022eclipse}.

Most recently \cite{ozdonmez2023investigation} added 36 new mid-eclipse times to the historical database and investigated four possible circumbinary models to explain the observed variations in O – C values.  They considered two datasets, the first included all recorded eclipse times whilst the second considered only eclipse times with uncertainties of less than 2 seconds, so excluding all secondary eclipses.  For each data set they considered both a one-circumbinary and a two-circumbinary system.  

Their one-circumbinary model comprised a planet of minimum mass $\mathrm{\sim}$9.5M\textsubscript{J} and an orbital period of $\mathrm{\sim}$20.1y with both datasets yielding very similar system parameters.  This model was found to have an orbital stability timeline of 10My.  Adding a second planet with a small,  $\mathrm{\sim}$6s, LTTE contribution improved the fit to the O - C residuals but orbital stability could only be maintained over a small range of system parameters for their restricted data set. The uncertainties of some observations recorded in the full data set are comparable to the magnitude of the LTTE of the second planet which, they believe, casts doubt on the validity of the orbital stability calculations made with the full data set.

\begin{figure}
	\captionsetup{width=8.50cm}
   \includegraphics[width=8.50cm]{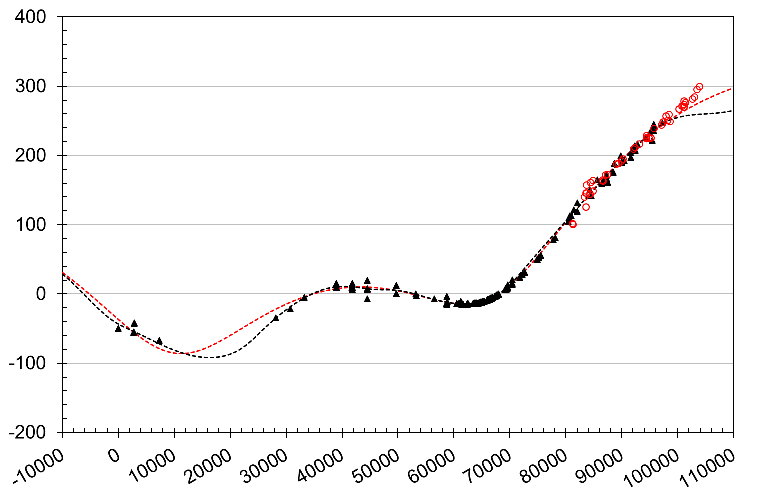}
     \caption{O - C chart for NN Ser showing the two models from \protect\cite{ozdonmez2023investigation} updated with new data from this paper. Historical data $\mathrm{\sim}$ black filled triangles; Our new data $\mathrm{\sim}$ red open circles; 1-planet model $\mathrm{\sim}$ red dashed line; 2-planet model $\mathrm{\sim}$ black dashed line; Vertical axis seconds; Horizontal axis cycle number.}
     \label{NN_Ser_Oz_models}
\end{figure}

The O - C residuals are shown in Fig. \ref{NN_Ser_Oz_models} where we include our 20 new times of minima made post 2022 March where it is seen that both models fail to correctly predict our new O - C values. The new data does not show the flattening of cyclical variation in vicinity of E $\mathrm{\sim}$100,000 predicted by these models and most notably for the 2 planet model.

With our new data we present an updated quadratic ephemeris, Eq. \ref{NN_Ser_ephem_quad}, where we note the quadratic term has a positive coefficient that cannot be explained by either gravitational radiation or magnetic breaking.  

   \begin{equation}\label{NN_Ser_ephem_quad}
   \begin{aligned}
      BTDB_{\mathrm{min}} ={} & 2447344.5252(1) + 0.130080080(3)E  + \\
      & 7.7(3)  \times 10^{-13}E^2
   \end{aligned}
   \end{equation}

\begin{figure}
	\captionsetup{width=8.50cm}
   \includegraphics[width=8.50cm]{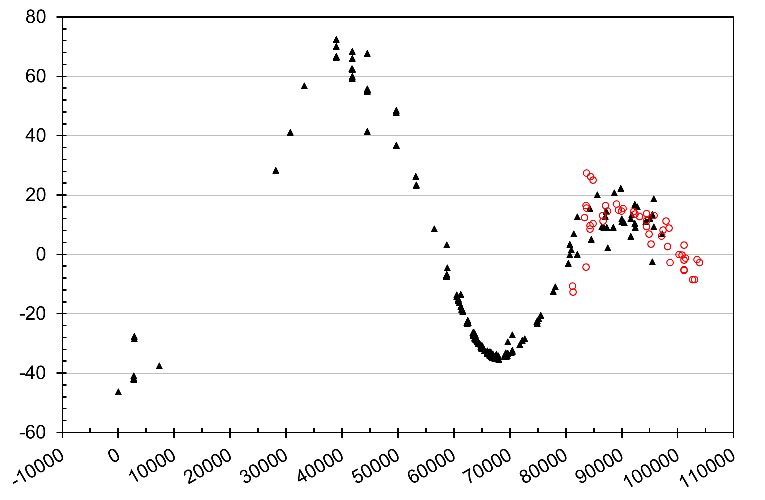}
     \caption{O - C chart for NN Ser showing the updated quadratic ephemeris in Eq. \ref{NN_Ser_ephem_quad}. Historical data $\mathrm{\sim}$ black filled triangles; Our new data $\mathrm{\sim}$ red open circles; Vertical axis seconds; Horizontal axis cycle number.}
     \label{NN_Ser_latest_ephem}
\end{figure}

The corresponding O - C chart is shown in Fig. \ref{NN_Ser_latest_ephem}.


\subsection{NSVS 07826147}
\cite{kelley2007combined} identified NSVS 07826147 as a potential sdB short period, $\mathrm{\sim}$3.9h, eclipsing binary which was subsequently confirmed by \cite{for2010rare} as an sdB+M5 PCEB. \cite{zhu2015circumbinary} were first to identify a small cyclical variation in the O - C eclipse timings having an approximate amplitude and period of 3.5s and 4.9y respectively.  They modelled this variation as a potential circumbinary planet of mass 4.7M\textsubscript{J} but details of its orbital parameters have not been released. For an overview of this binary system see \cite{pulley2018quest}.

\cite{lee2017long}, provided a further 59 primary and 52 secondary eclipse times and with 12 years of data gathered between 2005 to 2016, their analysis confirmed that this is a detached system with an sdB primary but with an M7 main sequence companion.  They found a 3.9s delay in the arrival time of the secondary eclipse which could be explained by a R{\o}mer delay and a possible small orbital eccentricity, but found no clear evidence to support the presence of a circumbinary third body.

\begin{figure}
	\captionsetup{width=8.50cm}
   \includegraphics[width=8.50cm]{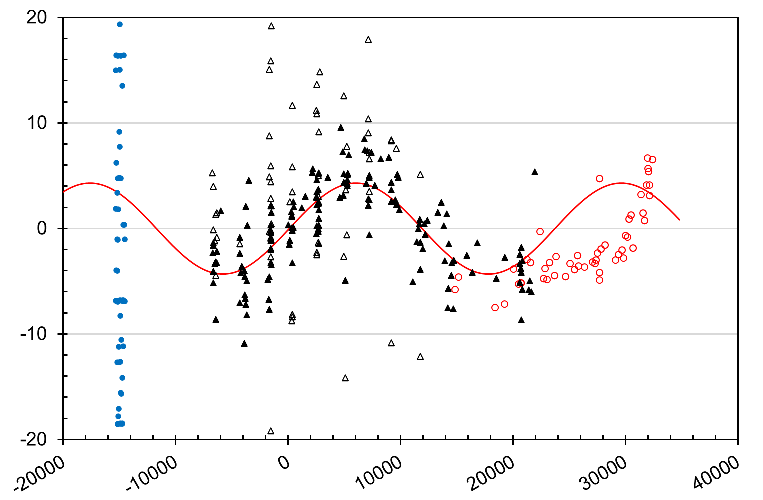}
     \caption{NSVS 07826147 O - C residuals for \protect\cite{wolf2021possible}). Model $\mathrm{\sim}$ red line; Primary eclipses are shown with solid triangles and open triangles are secondary eclipses. Part of SuperWASP unbinned data $\mathrm{\sim}$ blue circles; Our new data $\mathrm{\sim}$ red circles. Note, \protect\cite{wolf2021possible} did not use the SuperWASP data or secondary minima in their analysis. Vertical axis seconds; Horizontal axis cycle number.}
     \label{NSVS07826147_Wolf}
\end{figure}

The most recent investigation, \cite{wolf2021possible}, recorded 46 new primary eclipse times including six from TESS and a further four secondary eclipse times.  Their analysis, which did not include either the SuperWASP data nor secondary minima due to their large scatter, suggested the presence of a cyclical variation in the eclipse times of amplitude 4.3s and period of 10.5y, significantly longer than suggested by \cite{zhu2015circumbinary}, and possibly indicating the presence of a 1.4M\textsubscript{J} circumbinary planet.  However our new data post 2021 March departs from their predicted model, see Fig. \ref{NSVS07826147_Wolf}

We present a further 51 new times of primary minima post 2017 September and from our analysis, excluding secondary minima for the above reasons, we determined the best fit ephemeris, Eq. \ref{NSVS07826147_ephem_quad}, with a negative quadratic coefficient:

   \begin{equation}\label{NSVS07826147_ephem_quad}
   \begin{aligned}
      BTDB_{\mathrm{min}} ={} & 2455611.926576(5) + 0.161770447(1)E - \\
      & 0.9(3)  \times 10^{-13}E^2
   \end{aligned}
   \end{equation}

\begin{figure}
	\captionsetup{width=8.50cm}
   \includegraphics[width=8.50cm]{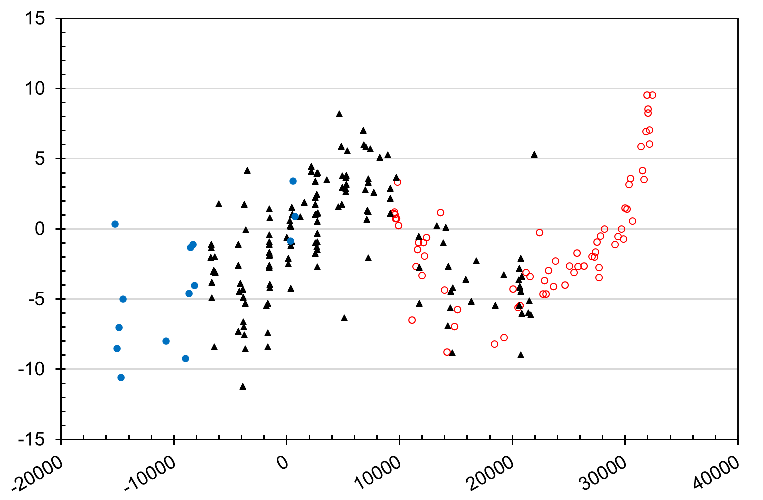}
     \caption{NSVS 07826147 O-C residuals calculated from Eq. \ref{NSVS07826147_ephem_quad}, all data are primary eclipses.  Binned SuperWASP data $\mathrm{\sim}$ blue circles; Historic data $\mathrm{\sim}$ black triangles; Our new data $\mathrm{\sim}$ red circles; Vertical axis seconds; Horizontal axis cycle number.}			  		
     \label{NSVS07826147_New_Ephem}
\end{figure}

The resulting O - C times calculated from this ephemeris include SuperWASP minima binned into 30 day periods, \cite{lohr2014period}, and shown in Fig. \ref{NSVS07826147_New_Ephem}.  It can be seen that with our new data, $E>30000$, the O - C values appear to be rapidly increasing.

\subsection{NSVS 14256825 (V1828 Aql)}
NSVS 14256825 was identified as a 13.2 magnitude variable star in the Northern Sky Variability Survey (NSVS), \cite{wozniak2004northern}, and subsequently as a short period, $\mathrm{\sim}$2.6h, eclipsing sdOB/dM binary by \cite{wils2007nsvs}.  \cite{pulley2022eclipse} reviewed four circumbinary models proposed prior to 2020 (\cite{beuermann2012aquest}, \cite{almeida2013two}, \cite{nasiroglu2017there} and \cite{zhu2019close}) and showed that with the addition of new times of minima, all models fail to correctly predict the times of eclipse within a few months of their publication.

\begin{figure}
	\captionsetup{width=8.50cm}
   \includegraphics[width=8.50cm]{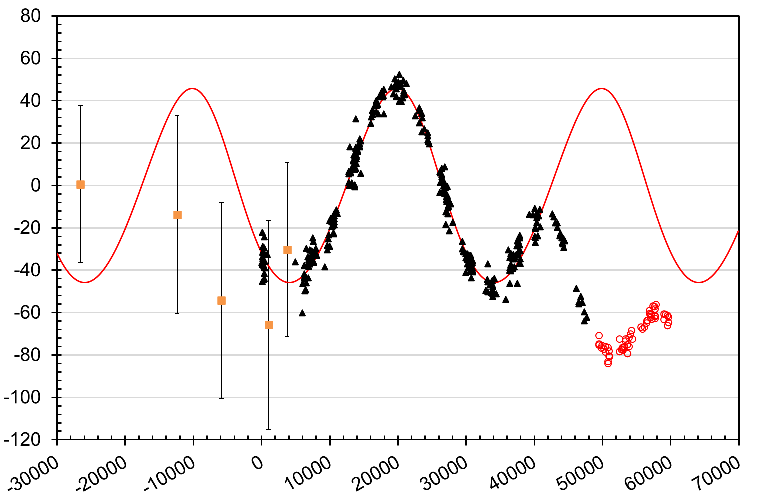}
     \caption{O – C chart for NSVS 14256825 from \protect\cite{nehir2022new} updated with new data from this paper. Historical data $\mathrm{\sim}$ black filled triangles; NSVS and ASAS data $\mathrm{\sim}$ orange squares; Our new data $\mathrm{\sim}$ red open circles; Circumbinary model $\mathrm{\sim}$ red line. Vertical axis seconds; Horizontal axis cycle number. Data departs from model in 2019 December, cycle $\mathrm{\sim}$ 41600}
     \label{NSVS14256825_Nehir}
\end{figure}

More recently \cite{nehir2022new} presented 16 new times of minima observed between 2018 July and 2019 August and derived updated parameters for this PCEB system.  Their O - C analysis embraces times of minima observed between 2007 and 2019 but excludes the earlier, less precise, NSVS and All Sky Automated Survey (ASAS) data. They found that the ETVs could be explained by the presence of a single, $\mathrm{\sim}$13M\textsubscript{J} planet with parameters similar to those obtained by \cite{zhu2019close}.  However, with new data post December 2019, it is seen that this model also fails to correctly predict new eclipse times, Fig. \ref{NSVS14256825_Nehir}, where we have included the NSVS and ASAS data for completeness.

\begin{figure}
	\captionsetup{width=8.50cm}
   \includegraphics[width=8.50cm]{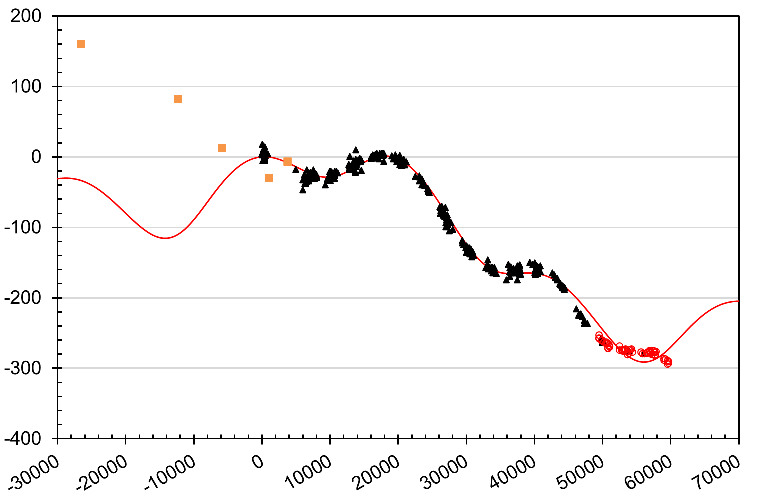}
     \caption{O – C chart for NSVS 14256825 showing the 2-body model from \protect\cite{zervas2024nsvs} updated with new data from this paper. Historical data $\mathrm{\sim}$ black filled triangles; NSVS and ASAS data $\mathrm{\sim}$ orange squares; Our new data $\mathrm{\sim}$ red open circles; Circumbinary model $\mathrm{\sim}$ red line. Vertical axis seconds; Horizontal axis cycle number. Zervas et al. note the poor data fit post 2020 December, cycle $\mathrm{\sim}$ 46000}
     \label{NSVS14256825_Zervas}
\end{figure}

A comprehensive analysis of ETVs for this system is presented by \cite{zervas2024nsvs} who explored a possible two-circumbinary model employing a complex grid search on two datasets. Dataset A included all recorded times of minima between 1999 to 2024 whilst dataset B excluded the less precise NSVS and ASAS datasets so covering the observational period 2007 to 2024. For both datasets their best fit to the O - C diagram yielded constrained masses for both planets but with orbits that were unstable on a timescale of a few hundred years.  However, they did find several hundred orbits stable for more than 1My within a 90\% confidence level of the best fitting curve but none of these solutions constrain the outer planet’s mass and period noting that the fit to the NSVS and ASAS data was poor and not perfect for data post 2021.  While they were unable to favour a single model, they concluded that the ETV data could be explained by the presence of an inner circumbinary planet of mass $\mathrm{\sim}$11M\textsubscript{J} in a nearly circular orbit of period $\mathrm{\sim}$7y. The outer planet is unconstrained with a period of between 20y and 50y and mass in the range of 11M\textsubscript{J} to 70M\textsubscript{J}.  We show one example O - C plot, Fig. \ref{NSVS14256825_Zervas}, where we include 56 new times of minima post 2022 April, confirming the imperfect fit to our new data 

\begin{figure}
	\captionsetup{width=8.50cm}
   \includegraphics[width=8.50cm]{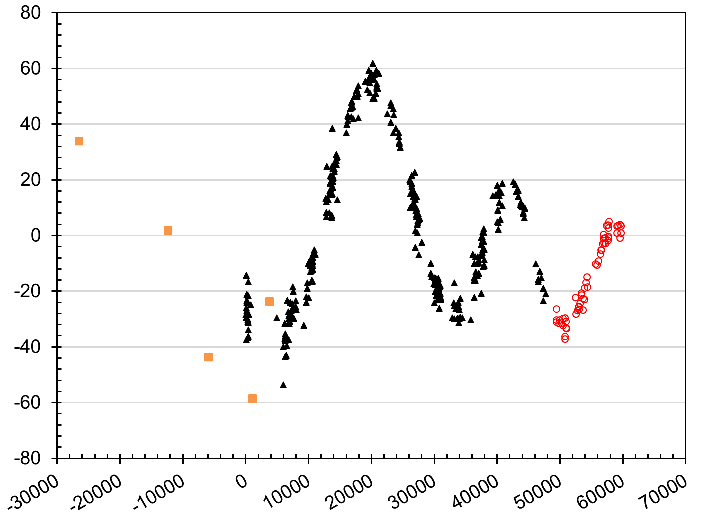}
     \caption{O – C chart for NSVS 14256825 with our quadratic ephemeris of Eq. \ref{NSVS_14256825_ephem_quad}. Historical data $\mathrm{\sim}$ black filled triangles; NSVS and ASAS data $\mathrm{\sim}$ orange squares; Our new data $\mathrm{\sim}$ red open circles; Vertical axis seconds; Horizontal axis cycle number.}
     \label{NSVS14256825_Quad}
\end{figure}

With all available data we present a revised quadratic ephemeris, Eq. \ref{NSVS_14256825_ephem_quad} and corresponding O-C diagram, Fig. \ref{NSVS14256825_Quad}.

   \begin{equation}\label{NSVS_14256825_ephem_quad}
   \begin{aligned}
      BTDB_{\mathrm{min}} ={} & 2454274.20913(4) + 0.1103741092(3)E - \\
      & 2.6(5) \times 10^{-13}E^2
   \end{aligned}
   \end{equation}

\subsection{NY Vir}
\begin{figure*}
\centering
   \includegraphics[width=17cm]{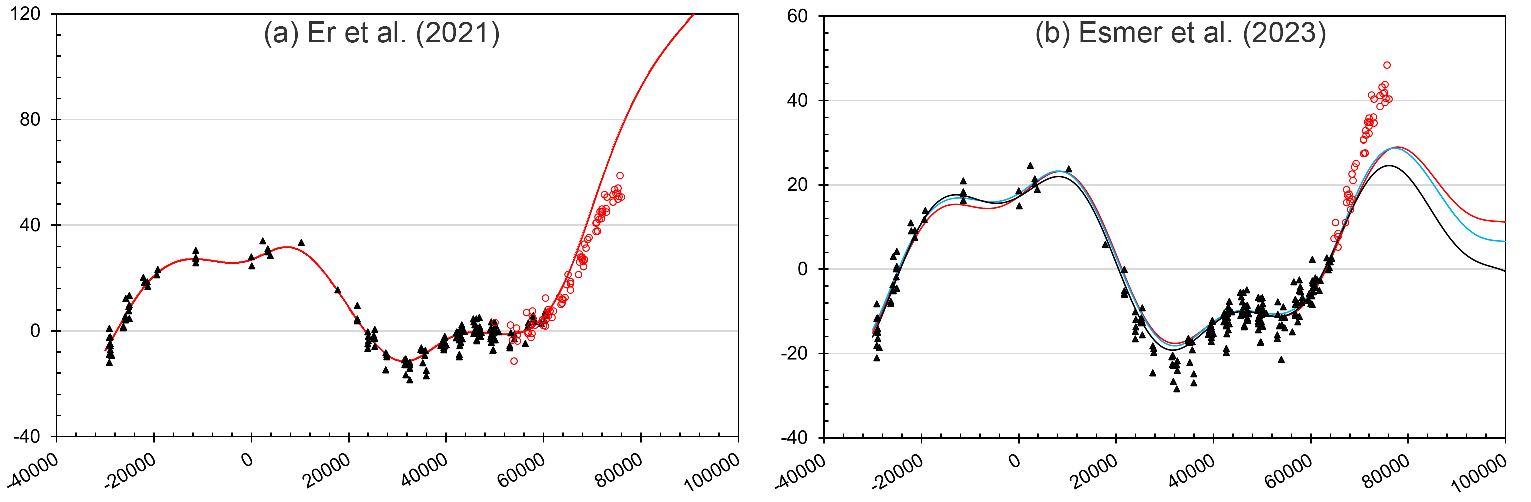}
     \caption{O - C diagrams for NY Vir; Historical data $\mathrm{\sim}$ black filled triangles; Our new data $\mathrm{\sim}$ red open circles; Vertical axis seconds; Horizontal axis cycle number. (a) \protect\cite{er2021new} model shown in red and beginning to show departure from data post 2023; (b) \protect\cite{esmer2023testing}) models red line linear ephemeris and zero eccentricity: blue line linear ephemeris and less than 0.05 eccentricity; black line quadratic ephemeris and zero eccentricity  }
     \label{NY_Vir_Charts}
\end{figure*}

NY Vir was first reported as a short period, $\mathrm{\sim}$2.4h, eclipsing sdB/M5 non-contact binary system \citep{kilkenny1998ec} and subsequent observations revealed quasi-sinusoidal variations in its eclipse times.  Investigations into the cause of these ETVs has led six authors to propose the presence of one or two circumbinary planets and five have been reviewed, \cite{pulley2022eclipse}, who found that four, (\cite{qian2012bcircumbinary}; \cite{lee2014pulsating}; \cite{bacsturk2018orbital}; \cite{song2019updated}), failed to predict future eclipse times with new observations.  The fifth proposal, \cite{er2021new}, indicated that the observed eclipse times post mid-2021, were occurring before that predicted by their model but more observations were necessary to confirm this trend.  However, with our new data collected through to 2025, it can be confirmed that the observed eclipses are now occurring some 17s before that predicted by their model, see Fig. \ref{NY_Vir_Charts}a, and in addition, \cite{mai2022orbital} have shown that the \cite{er2021new} model is unstable on timescales of $\mathrm{\sim}$3My, much less than the presumed lifetime of the PCEB phase.

Most recently a sixth group of models has been proposed by \cite{esmer2023testing} who considered a two-circumbinary system with both a linear and quadratic ephemeris. They found solutions with a quadratic ephemeris tended to have eccentric planetary orbits and more likely to be unstable leading them to prefer solutions with small or zero eccentricities.  The O-C diagram for their three preferred solutions, linear ephemeris with small eccentricities and linear and quadratic ephemerides with zero eccentricities, are shown in Fig. \ref{NY_Vir_Charts}b.  We have included our new data post 2022 May which confirms that these models fail to predict correctly new eclipse times.

We present 47 new times of minima and with all available data, but excluding TESS data, we compute a revised quadratic ephemeris with a positive quadratic coefficient, Eq. \ref{NY_Vir_ephem_quad}:

\begin{equation}\label{NY_Vir_ephem_quad}
   \begin{aligned}
      BTDB_{\mathrm{min}} ={} & 2453174.44278(2) + 0.1010159602(5)E + \\
      & 1.4(1) \times 10^{-13}E^2
   \end{aligned}
\end{equation}

\begin{figure}
	\captionsetup{width=8.50cm}
   \includegraphics[width=8.50cm]{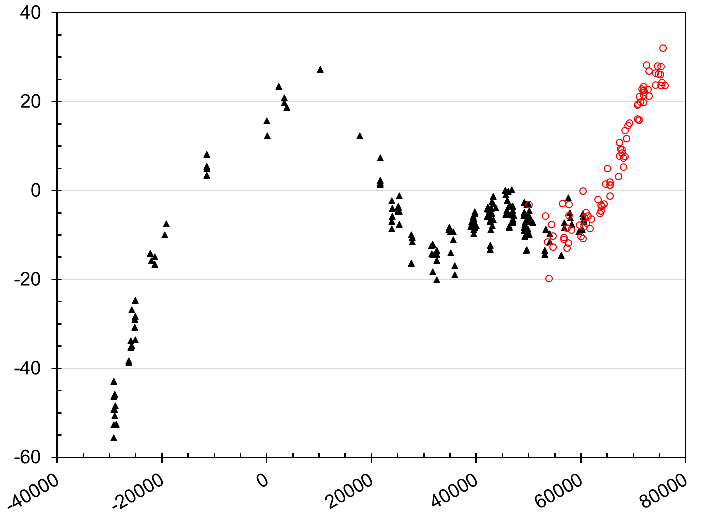}
     \caption{O – C chart for NY Vir with our quadratic ephemeris of Eq. \ref{NY_Vir_ephem_quad}. Historical data $\mathrm{\sim}$ black filled triangles; Our new data $\mathrm{\sim}$ red open circles; Vertical axis seconds; Horizontal axis cycle number.}
     \label{NY_Vir_Quad_Chart}
\end{figure}

The O-C diagram for this ephemeris is shown in Fig. \ref{NY_Vir_Quad_Chart} indicating the ETVs as having a possible quasi-periodic form.

\subsection{QS Vir (EC13471-1258)}
QS Vir was discovered in the Edinburgh-Cape faint blue object survey and subsequently identified as a short period, $\mathrm{\sim}$3.6 hours, eclipsing binary comprising a DA white dwarf with a subdwarf, M3.5-M4.5, flaring companion, \cite{o2003da+}.  The subdwarf is believed to almost fill its Roche Lobe leading the authors to categorise it as a hibernating cataclysmic variable but subsequently \cite{parsons2011stellar} suggested QS Vir could be a pre-cataclysmic variable. Between 1992 and 2016, 110 times of minima had been reported and in this paper we note a further 34 mid-eclipse times we observed post 2019.

\begin{figure*}
\centering
   \includegraphics[width=17cm]{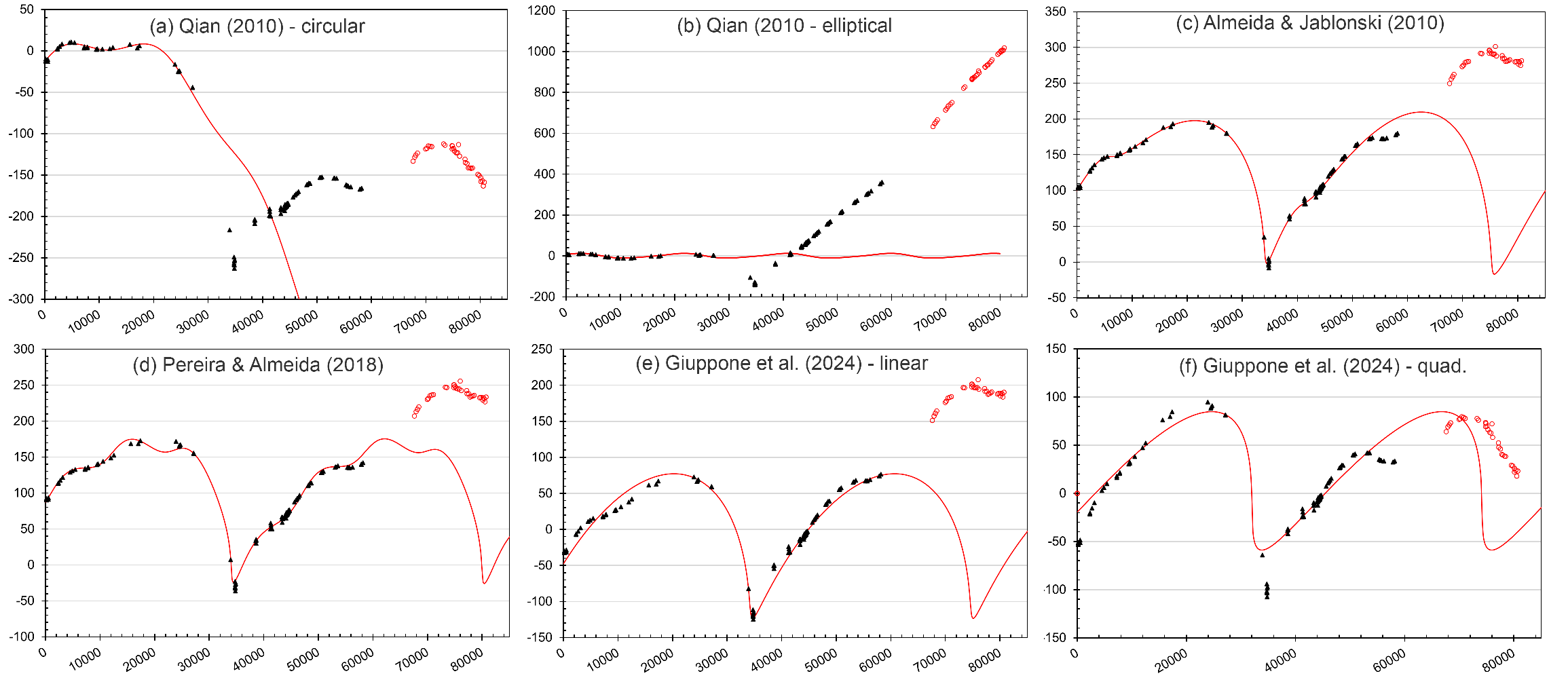}
     \caption{O--C plots for QS Vir.  Vertical axis is seconds and horizontal axis cycle number.  Historical data $\mathrm{\sim}$ black solid triangles; our new data $\mathrm{\sim}$ red open circles; circumbinary model $\mathrm{\sim}$ red line. }
     \label{QS_Vir_Charts}
\end{figure*}

Variations in the orbital period of QS Vir were first reported by \cite{qian2010giant}) who attributed them to a giant circumbinary planet producing two models with the planet (i) in a circular orbit and (ii) in a 0.37 elliptical orbit.  Both models quickly failed to predict new observations, Figs \ref{QS_Vir_Charts}a and \ref{QS_Vir_Charts}b.  Incorporating additional times of minima, not available to Qian, which emphasised a large change in O-C values in the vicinity of E$\mathrm{\sim}$33000, \cite{parsons2010orbital} proposed a single circumbinary planet in a highly elliptical orbit.  Although not able to show the period variations were caused by other mechanisms, they were unsure of the validity of their circumbinary model and did not publish parameters.

\cite{almeidajablonski2010} also addressed the rapid change of eclipse times but with a two-circumbinary model, Fig. \ref{QS_Vir_Charts}c, with both planets in highly elliptical orbits, 0.62 and 0.92. However orbital stability analysis, \cite{horner2013detailed}, revealed this solution to be extremely dynamically unstable and, within 5 years, their model failed to predict new eclipse times, Fig \ref{QS_Vir_Charts}c.\footnote[8]{\cite{almeidajablonski2010}, {Table 1 defines the date of periastron for the $\zeta$4 term as the date of the reference epoch. For the best model fit we found we had to modify this date to 2441425}} 

A fourth circumbinary model was proposed by \cite{pereiraalmeida2018} consisting of two circumbinary bodies, a brown dwarf in a slightly eccentric orbit, 0.1, and a super Jupiter planet in a highly eccentric, >0.9, orbit.  When we add in our new data we find that this model also fails to predict eclipse times, Fig. \ref{QS_Vir_Charts}d, sometime between 2016 March and 2020 January.

Most recently \cite{giuppone2024unveiling}, using existing historical data through to 2016, proposed the presence of a low mass stellar third body in a highly eccentric orbit adopting first a linear and then a quadratic ephemeris.  Their linear ephemeris provided the best fit but they suggest a second circumbinary body is needed to improve the quadratic fit.  However we find that with the addition of our 34 new mid-eclipse times post 2019, neither model can adequately predict future eclipses, Fig. \ref{QS_Vir_Charts}e and Fig. \ref{QS_Vir_Charts}f.

\begin{figure}
	\captionsetup{width=8.50cm}
   \includegraphics[width=8.50cm]{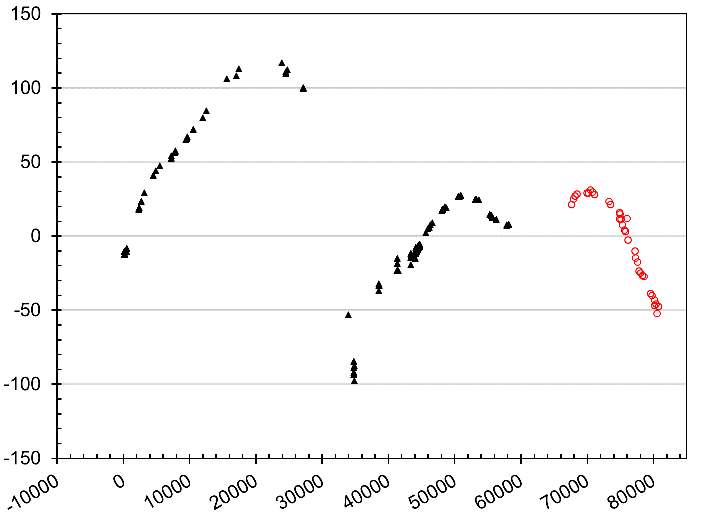}
     \caption{O – C chart for QS Vir showing our updated quadratic ephemeris in Eq. \ref{QS_Vir_ephem_quad}. Historical data $\mathrm{\sim}$ black filled triangles; Our new data $\mathrm{\sim}$ red open circles. Vertical axis is seconds and horizontal axis cycle number.}
     \label{QS_Vir_Quad_Ephem}
\end{figure}

With our new data the observational baseline is extended to 33 years giving a revised quadratic ephemeris as:

   \begin{equation}\label{QS_Vir_ephem_quad}
   \begin{aligned}
      BTDB_{\mathrm{min}} ={} & 2448689.6413(1) + 0.150757442(6)E - \\
      & 8.6(6) \times 10^{-13}E^2
   \end{aligned}
   \end{equation}

The O-C residuals calculated from Eq. \ref{QS_Vir_ephem_quad} are shown in Fig. \ref{QS_Vir_Quad_Ephem} where it can be seen that QS Vir exhibits amongst the largest O-C variations, exceeding 200 seconds, for these short period binary systems. 

\subsection{RR Cae} \label{rr_cae_section}
RR Cae was identified as a short period, $\mathrm{\sim}$7.3 h, eclipsing binary by \cite{krzeminski1984lft} comprising a cool DA white dwarf primary with a dwarf M4 main sequence companion, see \cite{pulley2022eclipse} for an overview of this system.  Times of eclipse minima have been recorded since 1984 April and variations in O-C of +/-13 secs led  \cite{qian2012acircumbinary} to propose the presence of a single circumbinary planet.  However, new data from \cite{bours2016long} revealed that within twelve months this planetary model would fail to predict future eclipse times.

\begin{figure}
	\captionsetup{width=8.50cm}
   \includegraphics[width=8.50cm]{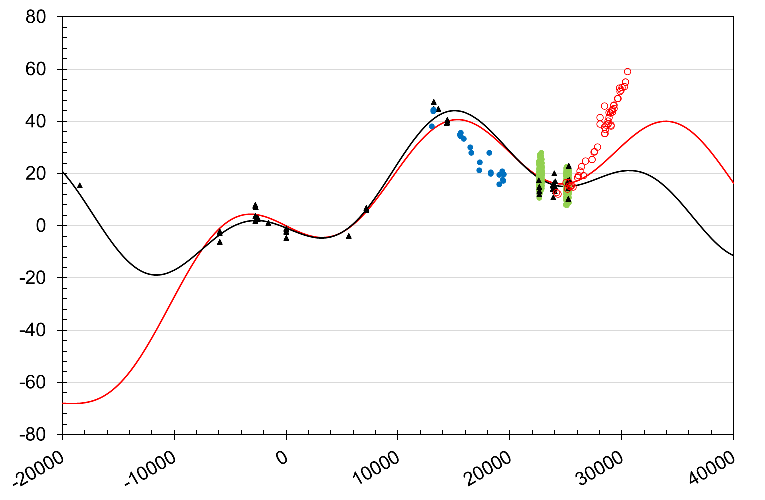}
     \caption{\protect\cite{rattanamala2023eclipse} models for RR Cae. 1 planet model $\mathrm{\sim}$ red line; 2 planet model $\mathrm{\sim}$ black line; Historic data $\mathrm{\sim}$ black triangles; Bours data $\mathrm{\sim}$ blue circles; TESS data from Rattanamala $\mathrm{\sim}$ green circles; Our new data $\mathrm{\sim}$ red open circles. Vertical axis seconds; Horizontal axis cycle number.}
     \label{RR_Cae_Ratt_plot}
\end{figure}

A second, and the most recent analysis of ETVs of this system, is presented by \cite{rattanamala2023eclipse} who proposed both a one-circumbinary and a two-circumbinary model.  The single-circumbinary model comprised a planet of minimum mass 3.4M\textsubscript{J} in a 16.9y orbit.  Their second model, which excludes \cite{krzeminski1984lft} data point, comprised two planets with minimum masses of 3.0M\textsubscript{J} and 2.7M\textsubscript{J} with orbital periods 15.0y and 39y respectively.  Both models are shown in Fig. \ref{RR_Cae_Ratt_plot} where we have included data from \cite{bours2016long}, not available to the authors, Rattanamala’s ground based observations and our unpublished data post 2019 December.  For completeness we have included the authors' TESS data (see Section \ref{ssec:tess}). and as can be seen from Fig. \ref{RR_Cae_Ratt_plot}, both models fail to predict our 35 new eclipse times of minima post 2021.

Excluding TESS data (see Section \ref{ssec:tess}), we compute a new quadratic ephemeris from the remaining times of minima:

   \begin{equation}\label{RR_Cae_ephem_quad}
   \begin{aligned}
      BTDB_{\mathrm{min}} ={} & 2451523.04870(3) + 0.303703648(3)E - \\
      & 1.1(1) \times 10^{-13}E^2
   \end{aligned}
   \end{equation}

\begin{figure}
	\captionsetup{width=8.50cm}
   \includegraphics[width=8.50cm]{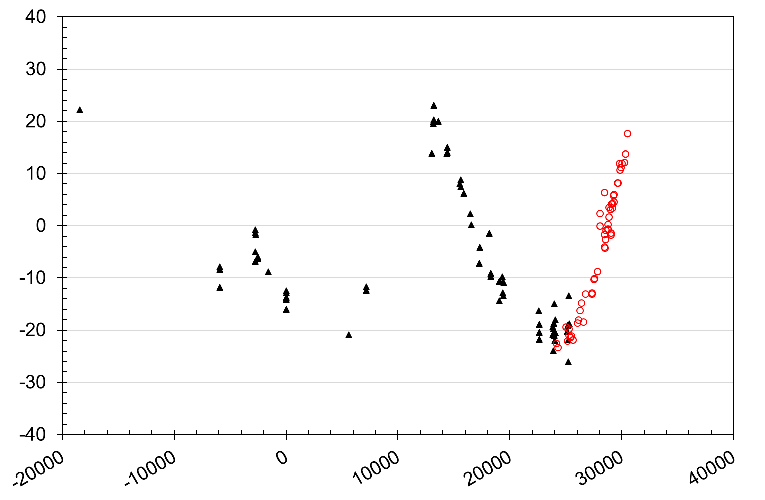}
     \caption{ Quadratic O-C residuals for RR Cae computed from Eq. \ref{RR_Cae_ephem_quad}. Historic data $\mathrm{\sim}$ black circles; Our new data $\mathrm{\sim}$ red circles. Vertical axis seconds; Horizontal axis cycle number.}
     \label{RR_Cae_our_new_ephem}
\end{figure}

With this ephemeris a revised O-C curve, Fig. \ref{RR_Cae_our_new_ephem}, shows apparent quasi-cyclical timing variations of the order of +/-20s but we anticipate that this ephemeris will be further refined as more data becomes available over the next two decades.

\section{ORIGINS of ETVs}
The observed variations in eclipse times of binary systems can be driven by a variety of astrophysical and dynamical phenomena (see for example \cite{zorotovic2013origin}, \cite{borkovits2025then} and references therein), including (a) stellar winds and magnetic breaking (b) gravitational radiation (c) apsidal motion (d) relativistic effects (e) the presence of circumbinary objects (f) magnetic activity within the secondary star, (g)  abrupt changes due to mass transfer and (h) star spots and  pulsations. These processes broadly fall into two categories: those giving rise to secular period changes and those that yield quasi-cyclical variations in eclipse times and frequently observed in short period PCEB systems.

\subsection{Secular period changes}
Long term continuous period changes can occur through angular momentum loss (AML) resulting in the binary period decreasing which is inferred in the system's ephemeris by the inclusion of a negative coefficient quadratic term.  Stellar winds interacting with the systems' magnetic fields can generate magnetic breaking and the consequential AML brings the binary pair closer together.  Similarly, as the binary pair orbit each other the emission of gravitational waves (gravitational radiation) carries energy away from the system causing the binary separation and period to decrease. Mass transfer from one star to the other will lead to a period change, although that is unlikely for detached systems such as the binaries studied in this paper. For RR Cae a model developed by \cite{rattanamala2023eclipse} proposed the existence of a cold star spot to fit their data, but as pointed out in Section \ref{rr_cae_section}, this model does not fit our recent data.

\subsection{Quasi-cyclical changes}
Phenomena most often included in this category are relativistic effects, apsidal motion, magnetic effects in the compact secondary star and the presence of one or more circumbinary objects.  Relativistic effects are generally small and cannot explain the magnitude of the observed ETVs. 
Apsidal motion is dependent upon the magnitude of the eccentricity of the binary orbit.  However, for these short period PCEB systems with the small separation between the binary components, it is expected the binary orbit will quickly circularise making the ETV contribution from apsidal motion minimal.  For further information see \cite{pulley2022eclipse} and references therein.  
Magnetic effects, e.g. \cite{applegate1992mechanism}, associated with the compact secondary star and the presence of circumbinary objects have been the most widely considered mechanisms for explaining these observed ETVs:

\subsubsection{Magnetic mechanisms}
The Applegate mechanism \citep{applegate1992mechanism} can contribute to the periodic modulation of the O - C curve for systems with a magnetic secondary star which changes its shape resulting in a change in quadrupole moment in a thin outer shell of the secondary. A significant improvement in this model was made by Völschow’s two-zone model \citep{volschow2016eclipsing} who considered the changes of quadrupole moments in the core and an outer shell of different densities. With the assumption that the two stars of the binary system are tidally locked, they derived an expression for the magnetic energy, $\Delta$E, required as a fraction of the energy, ${E_{sec}}$, radiated by the secondary during the period of the modulation, Eq. \ref{mag_volschow_eq_35}.

\begin{equation}\label{mag_volschow_eq_35}
\frac{\Delta E}{E_{sec}} \; = \; k_1 \; . \; \frac{M_{sec}R^2_{sec}}{P^2_{bin}P_{mod}L_{sec}} \; . \; \left( 1 \pm \sqrt{1 - k_2 G \frac{a^2_{bin} M_{sec} P^2_{bin} }{R^5_{sec}} \frac{\Delta P}{P_{bin}} } \right)^2
\end{equation}

\noindent where the value of $k1 = 0.133$ and $k2=3.42$ for low-mass stars, $M_{sec}$, $R_{sec}$ and $L_{sec}$ are the mass, radius and luminosity of the secondary respectively, $P_{bin}$ and $a_{bin}$ are the binary period and the semi-major axis of the binary, $P_{mod}$ is the period of the modulation and $\Delta P / P_{bin}$ is the fractional period change equal to $4 {\pi} K / P_{mod}$ where $K$ is the semi-amplitude of the O - C modulation. $G$ is the gravitational constant.

Table \ref{table_magnetic_1} lists the system parameters for the seven binaries in this study. For the most recent models published, Table \ref{table_magnetic_2} lists the values of the modulation period  $P_{mod}$ and the semi-amplitude of the modulation, $K$, together with the values of $\Delta E / E_{sec}$ from Eq. \ref{mag_volschow_eq_35}. We have assumed that any of the modulations listed are potentially due to magnetic effects. However, the values of the ratios $\Delta E / E_{sec}$ are seen to be much larger than 1 in all cases. 

A different magnetic mechanism for non-tidally locked binaries was proposed in a simple model by \cite{lanza2020internal} . Spin-orbit coupling results in a varying quadrupole moment due to a non-axisymmetric magnetic field in the conduction zone of the secondary. The resulting change in rotational energy, $\Delta E_{rot}$, as a fraction of the radiant energy available, $E_{sec} = L_{sec} P_{mod}$, is given by (see \citeauthor{pulley2022eclipse} \citeyear{pulley2022eclipse} Eqs. 9 and 10)

\begin{equation}\label{mag_pulley_eq}
\frac{\Delta E_{rot}}{E_{sec}} = \frac{1}{3} \frac{m a^2_{bin}}{L_{sec}P_{mod}} \left( \frac{\Delta P}{P_{bin}} \right) \left(\frac{2\pi}{P_{bin}} \right)^2
\end{equation}

\noindent where $m$ is the reduced mass of the binary.
\vskip 0.20cm
Table \ref{table_magnetic_2} shows the values of the ratio $\Delta E_{rot} / E_{sec}$  for the \cite{lanza2020internal} model. Again, all the ratios are greater than 1, except for the second modulation in Rattanamala’s 2-planet model for RR Cae \citep{rattanamala2023eclipse}, where $\Delta E_{rot} / E_{sec} = 0.7$. Hence the Lanza mechanism is possibly contributing to the modulation in RR Cae only. 

\begin{table*}
\caption{System parameters for the seven binaries in this study.}
\label{table_magnetic_1}
\begin{flushleft}
\begin{tabular}{p{2.75cm} | c | c | c | c | c | c | c | l}
\hlineB{2.0}
Binary system & $P\textsubscript{bin}$ (d) & $a\textsubscript{bin}$/$R_{\sun}$ & $M_{pri}$/$M_{\sun}$ & $M_{sec}$/$M_{\sun}$ & $R_{sec}$/$R_{\sun}$ & $T_{sec}$ (K) & $L_{sec}$/$L_{\sun}$ & Refs.\\
\hlineB{2.0}
HS0705+6700 & 0.09564 & 0.75 & 0.483 & 0.134 & 0.186 & 2900 & 0.00219 & 1,2,a\\
NN Ser & 0.13008 & 0.934 & 0.535 & 0.111 & 0.149 & 2920 & 0.00147 & 3\\
NSVS 07826147 & 0.16177 & 0.98 & 0.376 & 0.113 & 0.152 & 3100 & 0.00189 & 8,b\\
NSVS 14256825 & 0.11037 & 0.74 & 0.351 & 0.095 & 0.14 & 2550 & 0.002 & 4\\
NY Vir & 0.101016 & 0.764 & 0.466 & 0.122 & 0.16 & 3000 & 0.00142 & 5,2\\
QS Vir & 0.15076 & 1.27 & 0.78 & 0.43 & 0.42 & 3100 & 0.01 & 6,7\\
RR Cae & 0.3037 & 1.62 & 0.453 & 0.168 & 0.209 & 3100 & 0.00362 & 10\\
\hlineB{2.0}
\end{tabular}
\end{flushleft}
\begin{flushleft}
\begin{minipage}{17cm}
\small \textbf{References:} (1) \cite{drechsel2001hs}; (2) \cite{volschow2016eclipsing}; (3) \cite{parsons2010precise}; (4) \cite{nehir2022new}; (5) \cite{vuvckovic2007binary}; (6) \cite{giuppone2024unveiling}; (7) \cite{o2003da+}; (8) \cite{wolf2021possible}; (9) \cite{for2010rare}; (10) \cite{rattanamala2023eclipse}; (11) \cite{Er_Özdönmez_Nasiroglu_Kenger_2024}; (12) \cite{er2021new}; (13) \cite{zervas2024nsvs}; (14) \cite{ozdonmez2023investigation}; (15) \cite{esmer2023testing}; (a) $a_{bin}/R_{\sun}$ from Kepler's 3rd law; (b) $L_{sec}/L_{\sun}$ from Stefan-Boltzmann law
\end{minipage}
\end{flushleft}
\end{table*}

\begin{table*}
\caption{Energy ratios calculated for magnetic effects for the systems in this study. (The reference numbers are shown below Table \ref{table_magnetic_1})}
\label{table_magnetic_2}
\begin{flushleft}
\begin{tabular}{p{2.75cm} | >{\centering}p{1.0cm} | c | c | >{\centering}p{1.1cm} | c | c | >{\centering}p{1.1cm} | >{\centering}p{1.1cm} | >{\centering}p{1.1cm} | l}
\hlineB{2.0}
Binary system & O -- C Mod. & K (s) & $P\textsubscript{mod}$ (yr) & $\Delta P$/$P\textsubscript{bin}$ $(x10\textsuperscript{-6})$ & $\Delta E$/$E\textsubscript{sec}$ & m/$M_{\sun}$ & $E\textsubscript{sec}$ (J) $(x10\textsuperscript{32})$ & $\Delta E\textsubscript{rot}$ (J) $(x10\textsuperscript{33})$ & $\Delta E\textsubscript{rot}$/$E\textsubscript{sec}$  & Refs.\\
\hlineB{2.0}
HS0705+6700 & 1 of 2 & 85.2 & 8.10 & 4.19 & 181.3 & 0.10490 & 2.18 & 45.86 & 210.2 & 11\\
HS0705+6700 & 2 of 2 & 76.7 & 9.60 & 3.18 & 86.7 & 0.10490 & 2.59 & 34.83 & 134.7 & 11\\
HS0705+6700 & 1 of 3 & 71.0 & 7.70 & 3.67 & 145.3 & 0.10490 & 2.07 & 40.20 & 193.8 & 11\\
HS0705+6700 & 2 of 3 & 73.7 & 13.30 & 2.21 & 29.6 & 0.10490 & 3.58 & 24.16 & 67.4 & 11\\
HS0705+6700 & 3 of 3 & 196.0 & 38.20 & 2.04 & 8.8 & 0.10490 & 10.29 & 22.37 & 21.7 & 11\\
NN Ser & 1 & 44.1 & 20.18 & 0.87 & 72.4 & 0.09193 & 3.65 & 7.00 & 19.2 & 14\\
NSVS 07826147 & 1 & 4.3 & 10.50 & 0.16 & 5.9 & 0.08689 & 2.44 & 0.88 & 3.6 & 8,9\\
NSVS 14256825 & 1 & 45.9 & 9.08 & 2.01 & 190.0 & 0.07476 & 2.23 & 11.48 & 51.4 & 4\\
NSVS 14256825 & 1 & 22.0 & 18.00 & 0.49 & 5.1 & 0.07476 & 4.43 & 2.78 & 6.3 & 13\\
NY Vir & 1 of 2 & 7.9 & 9.00 & 0.35 & 14.7 & 0.0967 & 1.57 & 3.28 & 20.9 & 12\\
NY Vir & 2 of 2 & 33.8 & 27.20 & 0.49 & 9.9 & 0.0967 & 4.75 & 4.65 & 9.8 & 12\\
NY Vir & 1 of 2 & 35.1 & 22.27 & 0.63 & 19.6 & 0.0967 & 3.89 & 5.89 & 15.2 & 15\\
NY Vir & 2 of 2 & 9.4 & 8.38 & 0.45 & 26.6 & 0.0967 & 1.46 & 4.19 & 28.6 & 15\\
QS Vir & 1 & 148.7 & 16.67 & 3.55 & 13.2 & 0.27719 & 20.50 & 118.55 & 57.8 & 6\\
RR Cae & 1 & 14.0 & 16.60 & 0.34 & 66.8 & 0.12255 & 7.39 & 1.99 & 2.7 & 10\\
RR Cae & 1 of 2 & 12.0 & 15.00 & 0.32 & 65.7 & 0.12255 & 6.68 & 1.89 & 2.8 & 10\\
RR Cae & 2 of 2 & 20.0 & 39.00 & 0.20 & 9.6 & 0.12255 & 17.36 & 1.21 & 0.7 & 10\\
\hlineB{2.0}
\end{tabular}
\end{flushleft}
\begin{flushleft}
\begin{minipage}{17cm}
\small \textbf{Notes:} O -- C Mod. stands for O -- C modulation, e.g. 1 for a 1-modulation model, 1 of 2 for modulation 1 of a 2-modulation model etc. 
\end{minipage}
\end{flushleft}
\end{table*}

\subsubsection{Circumbinary companions}
Over the past two decades variations in eclipse times of many short period PCEB’s have been recorded and these ETVs have been attributed to the presence of one or more circumbinary objects.  In this paper we have reviewed 18 circumbinary models from the seven PCEB’s listed in Table \ref{table_objects_summary} and find that 15 of these models fail to accurately predict future eclipse times as new data becomes available.  These findings are consistent with our earlier investigations (see \citeauthor{pulley2018quest} \citeyear{pulley2018quest}  and \citeauthor{pulley2022eclipse} \citeyear{pulley2022eclipse}).  With this work we have now investigated 49 published circumbinary models relating to 10 PCEB systems which includes the seven systems reported here together with DE CVn, HS2231+2441 and HW Vir. Of the 49 models investigated, 46 fail to predict future eclipse times whilst the three more recent models, NN Ser one-planet model, Fig. \ref{NN_Ser_Oz_models}, NSVS 14256825, Fig. \ref{NSVS14256825_Zervas} and NY Vir, Fig. \ref{NY_Vir_Charts}a, are showing indications of a deviation/poor fit to their respective models but more observations are needed to confirm this.

\begin{figure}
	\captionsetup{width=8.50cm}
   \includegraphics[width=8.50cm]{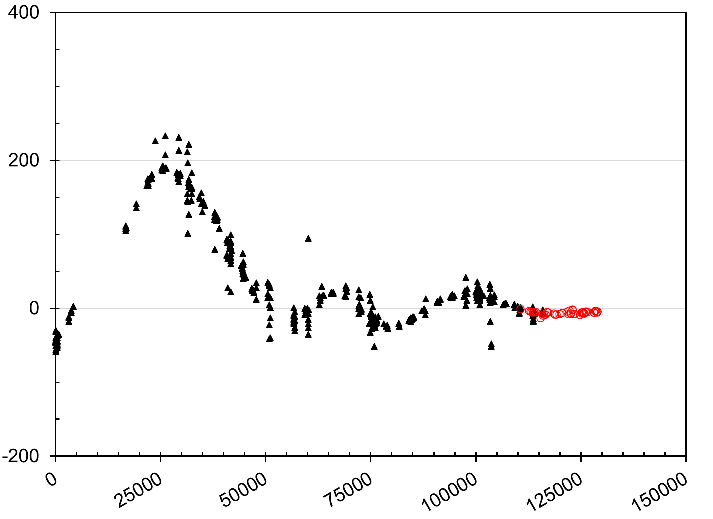}
     \caption{O -- C residuals for HW Vir from the best fit quadratic ephemeris indicating that the ETVs over recent years has diminished. Vertical axis seconds; Horizontal axis cycle number.}
     \label{HW_Vir_Quad_Ephem}
\end{figure}

O -- C plots based solely on the quadratic ephemeris present interesting comparisons between the various systems.  Many show strong activity with variations of between +/-20s and +/-100s, see for example Figs. \ref{NN_Ser_latest_ephem}, \ref{QS_Vir_Quad_Ephem} and \ref{RR_Cae_our_new_ephem} for NN Ser, QS Vir and RR Cae respectively.  However other systems show a diminishing level of activity over time for example HS0705+6700, Fig. \ref{HS0705_Quad_Ephem_graph}, and the prototype sdB eclipsing binary HW Vir, Fig. \ref{HW_Vir_Quad_Ephem}.  If these systems are approaching a quiescent phase in their ETVs it raises important questions on whether circumbinary objects are the major cause of these observed timing variations.

We also note that for several of these systems (HS0705+6700, NN Ser and NY Vir) the best fit ephemeris includes a quadratic term with a positive coefficient in the range $1.4 \times 10^{-13}$ to $7.7 \times 10^{-13}$.  There is no clear explanation as to the origins of the positive coefficient and historically the default explanation has always been the presence of an undetected very long period circumbinary body.  However, now with many years of observations, eclipse cycles recorded are approaching 100,000 and we can model the residual $O - C$, $(O - C)_{res}$, as follows:

\begin{equation}\label{oc_model_1}
(O - C)_{res} \; = \; T_m - T_c \; = \; T_m - \left(T_0 + PE + \beta E^2 + f(t) \right)
\end{equation}

\noindent where $T_m$ is the observed time of minimum at cycle $E$, $T_c$ is the calculated time of minimum, $T_0$ is the reference epoch, $P$ is the binary period, $\beta$ is the quadratic coefficient, which can be negative, positive or zero and assumed constant over the duration of the observations and $f(t)$ the eclipse time varying component and frequently expressed as a series of periodic terms. Thus Eq. \ref{oc_model_1} can be re-arranged to give:

\begin{equation}\label{oc_model_2}
\frac{T_m- T_0}{E} \; = \; P + \beta E + \frac{f(t)}{E} + \frac{(O - C)_{res}}{E}
\end{equation}

\noindent As the magnitude of $E$ increases the contribution from the two right hand terms become less significant and for large values of $E$, when $\beta E^2$ is greater than the amplitude of the periodic term $f(t)$, Eq. \ref{oc_model_2} approximates to:

\begin{equation}\label{oc_model_3}
\frac{T_m- T_0}{E} \; = \; P + \beta E
\end{equation}

\begin{figure}
	\captionsetup{width=8.50cm}
   \includegraphics[width=8.50cm]{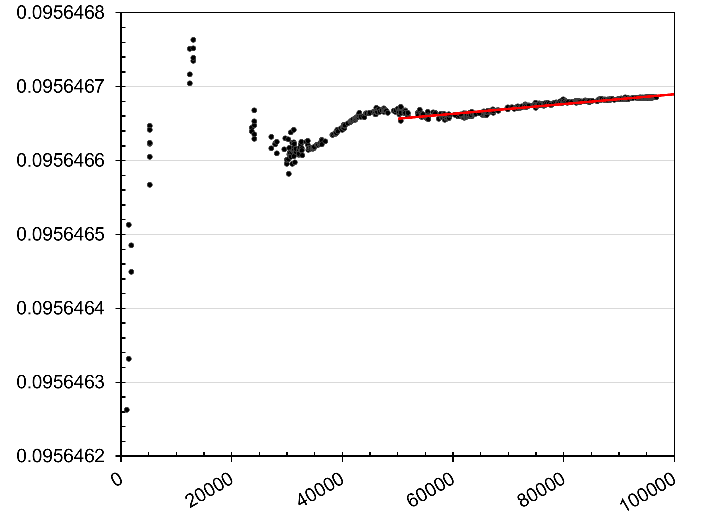}
     \caption{Time averaged binary period for HS0705+6700 indicating the presence of a positive quadratic coefficient, red trend line, with magnitude between $10^{-12}$d and $10^{-13}$d.  Horizontal axis, cycle number. (E); vertical axis, binary period in days.}
     \label{HS0705_Ser_Time_Avg_Period}
\end{figure}

\noindent As an example we show for HS0705+6700, Fig. \ref{HS0705_Ser_Time_Avg_Period}, a plot of $((T_m - T_0)/E)$ where it can be seen for large values of $E$ (>60,000) where the contribution from possible circumbinary companion(s) is small, the slope of the curve is positive, implying the quadratic coefficient is positive.

\section{Conclusions}
In this paper we report on 18 recent circumbinary models proposed for seven short period PCEB systems and find that 15 of these models clearly fail to predict our new eclipse times and three, NN Ser, NSVS 14256825 and NY Vir, show deviations from the ETV models.  In total, and over the past 25 years, these seven systems have had in excess of 42 circumbinary models put forward to explain their observed ETVs with many, e.g. HS0705+6700, having numerous "improved" models proposed as new data became available.  In all cases these proposed models have been short lived, failing to accurately predict new eclipse times of minima.  

Whilst ETV modelling based on circumbinary planets alone has yet to succeed in producing robust long-term eclipse time predictions for PCEBs, other proposed ETV explanations have also failed. Apsidal motion is an unlikely contender because the binary orbit will quickly circularise due to the close proximity of its two stellar components.  Furthermore, observations of secondary eclipses show these to occur at, or very close to, phase 0.5 i.e. midway between two primary eclipses. AML also provides another mechanism for binary period change through gravitational radiation or magnetic breaking but both mechanisms result in a unidirectional period decrease and cannot cause the quasi-cyclical ETVs seen in these systems.

Magnetic effects observed in the main sequence secondary star of the binary can also give rise to ETVs providing a strong contender to the circumbinary model.  However, the magnetic energy required to explain the observed O - C modulations is far too high to be a likely mechanism for the systems studied here, on the basis of the two magnetic models discussed (\citeauthor{volschow2016eclipsing} \citeyear{volschow2016eclipsing} and \citeauthor{lanza2020internal} \citeyear{lanza2020internal}), except in the case of the Lanza model for RR Cae. Given the approximations made in these models this may not be surprising, and further progress will require more complex models, such as the magnetohydrodynamic simulation of the stellar convection zone, used by \cite{navarrete2022origin}. Furthermore, the required magnetic energies are calculated, based on past models for the O - C modulations which our new data has shown to be incorrect by varying amounts. New models may therefore show significantly different values of the magnetic energy.

In addition, optimising the quadratic ephemeris frequently leads to a positive coefficient for the quadratic term.  There is currently no explanation for this and the original hypothesis of an undetected long period circumbinary planet looks more and more unlikely as new data extends the time line beyond two decades.  Furthermore, the optimised O - C plots for many of these systems show them appearing to enter a quiescent mode, e.g. HW Vir and HS0705+6700, whilst some appear to be moving from dormancy to energetic mode. It is possible to model these O - C plots by adding more periodic terms (planets?) but we have to question whether this is a realistic approach.

Whilst ETV analysis has yet to succeed in producing reliable circumbinary planetary models for these short period binary systems, it cannot be said that circumbinary planets do not exist orbiting these systems. Where they do exist, they could be generating only part of the observed ETVs with the remainder originating from other processes such as magnetic effects. However, to use the ETV methodology to predict the presence of circumbinary planets in these systems, it will be necessary to first remove these other causes and with the present level of understanding, particularly of magnetic effects, this may prove to be a difficult task.

\section*{Acknowledgements}
This work makes use of observations from the LCO network of telescopes and of the APASS data base maintained on the AAVSO website. This research has made use of the NASA Exoplanet Archive, which is operated by the California Institute of Technology, under contract with the National Aeronautics and Space Administration under the Exoplanet Exploration Program.  We extend our gratitude to Dr. Madelon Bours who made her observational data available to us for this work.  We like to thank Professor Robert Mutel (University of Iowa) for granting access to the Gemini telescope and Professor John Cannon (Macalester College \& MACRO) for granting access to the Robert L. Mutel telescope.  We thank Paul Downing (BAA) for two observations on NSVS 14256825. 
We gratefully acknowledge the financial support for this research project from The Royal Astronomical Society which helped fund the commercial remote telescope time required on targets not reachable from our own equipment. We would also like to thank Dr. Supachai Awiphan (National Astronomical Research Institute of Thailand), Professor  H{\"u}seyin Er (Atat{\"u}rk University), Dr. Ekrem Esmer (University of Ankara), Dr. Tony Lynas-Gray (University of Oxford), Professor Ilham Nasiro\v{g}lu (Atat{\"u}rk University), Aykut {\"O}zd{\"o}nmez (Atat{\"u}rk University) and Dr. Konstantinos Zervas (University of Patras) who addressed many of the questions we posed. We would also like to thank the referee whose guidance was greatly appreciated. Lastly we remember our colleague Americo Watkins who sadly passed away in 2025 June.


\section*{Data Availability}
The data underlying this article is available as on-line supplementary material.



\bibliographystyle{mnras}
\bibliography{altair} 



\appendix

\onecolumn

\section{Table of Observations}

\setlength{\LTleft}{12pt}

\begin{longtable}{lllllc}
\caption{ A sample of the compilation of our new times of minima observed between 2017 September and 2025 June. Telescopes used are listed in Table \ref{table_A2}. Full datasets are available as on-line supplementary material.}\\
\toprule
\multicolumn{1}{p{4.43em}}{Object} & \multicolumn{1}{p{6.855em}}{BJD\newline{}-2450000} & \multicolumn{1}{p{6.43em}}{Cycle} & \multicolumn{1}{p{4.43em}}{Uncertainty\newline{}(d)} & Filter & Telescope \\
    \midrule
HS0705+6700 & 9590.419089 & 81212.0 & 0.000118 & IRB & 3 \\
HS0705+6700 & 9590.514793 & 81213.0 & 0.000096 & IRB & 3 \\
HS0705+6700 & 9725.472158 & 82624.0 & 0.000025 & IRB & 1 \\
HS0705+6700 & 9811.554143 & 83524.0 & 0.000045 & IRB & 1 \\
HS0705+6700 & 9838.431051 & 83805.0 & 0.000115 & Clear & 2 \\
... & & & & & \\
RR Cae & 9813.247029 & 27297.0 & 0.0000347 & IRB & 9 \\
RR Cae & 9814.158142 & 27300.0 & 0.0000231 & IRB & 9 \\
RR Cae & 9880.061863 & 27517.0 & 0.0000243 & IRB & 10 \\
RR Cae & 9880.972976 & 27520.0 & 0.0000100 & IRB & 12 \\
    \bottomrule
  \label{table_A1}%
\end{longtable}%

\setlength{\LTleft}{12pt}

\begin{longtable}{cl}
  \caption{The Telescopes used in our observations. The values in the Ident column are cross referenced in the Telescope column in Table \ref{table_A1} above. }\\
    \toprule
    \textbf{Ident} & \multicolumn{1}{c}{\textbf{Telescope}} \\
    \midrule
1 & 235mm, Ham Observatory West Sussex, UK \\
2 & 203mm, Woodlands Observatory West Sussex, UK \\
3 & 104mm OG, PixelSkies, Castillejar, Spain \\
4 & 280mm, PixelSkies, Castillejar, Spain \\
5 & 508mm Gemini University of Iowa \\
6 & 508mm Warrumbungle Observatory, Siding Spring \\
7 & 432mm T21 iTelescope, Utah \\
8 & 508mm T11 iTelescope, Utah \\
9 & 508mm T30 iTelescope, Siding Spring \\
10 & 432mm T32 iTelescopes Siding Spring \\
11 & 432mm T17 iTelescope, Siding Spring \\
12 & 432mm Dubbo Observatory, NSW, Aus \\
13 & 305mm T18 iTelescope, Nerpio \\
14 & 2m Faulkes, Siding Spring \\
15 & 2m Faulkes, Hawaii \\
16 & 1m Faulkes, Siding Spring \\
17 & 1m Faulkes, McDonald Observatory \\
18 & 406mm Observatorio de La Divisa, Spain \\
19 & 432mm T19 iTelescope, Mayhill, NM \\
20 & 432mm BAA Alnitak, Spain \\
21 & 280mm Greenmoor, Oxfordshire, UK \\
22 & 365mm Astrognosis Observatory, MPO K01 \\
23 & 508mm Dubbo Observatory, NSW, Aus \\
24 & 432mm T19 iTelescope, Utah \\
25 & 1m Faulkes Mt Teide Tenerife \\
26 & 1m Faulkes Cerro Tololo, Chile \\
27 & 410mm Dubbo, SSON, Australia \\    \bottomrule
  \label{table_A2}%
\end{longtable}%

\twocolumn


\bsp	
\label{lastpage}
\end{document}